\documentclass[12pt, reqno]{amsart}
\usepackage{amsmath, amsthm, amscd, amsfonts, amssymb, graphicx, color}
\usepackage[bookmarksnumbered, colorlinks, plainpages]{hyperref}
\hypersetup{colorlinks=true,linkcolor=red, anchorcolor=green, citecolor=cyan, urlcolor=red, filecolor=magenta, pdftoolbar=true}

\usepackage{xcolor}
\usepackage{tikz}
\usepackage[onehalfspacing]{setspace}%
\theoremstyle{definition}

\theoremstyle{remark}

\numberwithin{equation}{section}

\begin{document}

\setcounter{page}{1}

\title[ The quest for the definition of life ]{ The quest for the definition of life  }
\author[Kazem Haghnejae Azar]{Kazem Haghnejae Azar}

\address{Mathematics Department, Faculty of Science, University of Mohaghegh Ardabili, Ardabil, Iran}
\email{\textcolor[rgb]{0.00,0.00,0.84}{haghnejad@uma.ac.ir}}

\subjclass[]{}


\begin{abstract} 
The intricacy and diversity inherent in living organisms present a formidable obstacle to the establishment of a universally accepted definition. Life manifests in a multitude of forms, exhibiting various attributes such as growth, reproduction, responsiveness to stimuli, adaptation, and homeostasis. However, each of these characteristics can also be observed to some degree within certain non-living systems, leading to a blurring of boundaries and generating conceptual complexities.
In this manuscript, I demonstrate that the transformation of a non-living entity into a living organism does not adhere to a specific temporal boundary that unequivocally designates the onset of life. Through mathematical analysis, I have demonstrated that a comprehensive definition of living beings does not exist, which means that there are no clear boundaries in the chemical processes that turn non-living entities into living ones. In other words, living organisms do not possess unique characteristics that can completely set them apart from non-living entities. Therefore, no definitive definition exists that unequivocally distinguishes living things from non-living things.

\end{abstract} \maketitle

\contentsline {section}{\numberline {1}Introduction}{}{section.1}
\contentsline {section}{\numberline {2}The quest for the definition of life}{}{section.2}
\contentsline {section}{\numberline {3}Mathematical method for examining the distinction between living and non-living organisms}{}{section.3}
\contentsline {section}{\numberline {4}Corollary}{}{section.4}
\contentsline {section}{\numberline {5}References}{}{section.5}

\section{Preliminaries and Introduction }
One prominent area of research focuses on understanding how the complex molecules necessary for life could have formed under the conditions present on early Earth. Many experiments have been conducted to simulate the environmental conditions of that time, such as the presence of reducing atmospheres, intense heat, lightning, and volcanic activity. These experiments have demonstrated that basic organic molecules, such as amino acids, nucleotides, and lipids, can be synthesized from simpler compounds.
Additionally, researchers have explored environments that might have provided suitable conditions for the emergence of life, such as hydrothermal vents on the ocean floor or clay-rich environments. These environments could have offered protection, a steady supply of energy and nutrients, and the necessary chemical reactions for the formation of complex biomolecules.
Delving into the differentiations between living and non-living entities necessitates a thorough comprehension of the intricate process by which a non-living chemical compound undergoes a metamorphosis into a living entity. Numerous inquiries have been raised by researchers concerning the origin of life and the transition from a non-living chemical state to a living one. For instance, in the emergence of life from prebiotic molecules approximately 3.5 billion years ago, what acted as the initial catalyst? By what mechanisms did prebiotic molecules acquire the ability to sustain and serve themselves? How might prebiotic molecules undergo purposeful transformations?

There are several existing definitions of living organisms as a phenomenon. Here are a few examples:
\begin{enumerate}
\item The classical definition: Living organisms are complex, self-replicating entities capable of growth, development, and responding to stimuli.
\item The cellular definition: Living organisms are composed of one or more cells, which are the basic structural and functional units of life.
\item The metabolic definition: Living organisms exhibit metabolic processes, such as the conversion of energy and the regulation of internal conditions.
\item The evolutionary definition: Living organisms are products of evolution, characterized by genetic variation, inheritance, and adaptation to their environment.
\item The homeostasis definition: Living organisms maintain a state of internal balance or homeostasis, regulating their internal conditions to survive.
\end{enumerate}
It's important to note that these definitions are not universally agreed upon and may vary depending on the scientific discipline or context in which they are used.

 The hypothesis of the transition from non-life to life, known as the "managed-metabolism" hypothesis, was first presented in detail by Stewart in (1918)\cite{24st}. The current complexity of life is the result of a multistep process that began with the catalytic chemistry of small molecules in simple networks, which are still preserved in the depths of metabolism. These reaction sequences were elaborated through processes of simple chemical selection, and only later took on the aspects of cellularization and organismal individuality (Kolb (2015)\cite{16k}).

The process of converting a non-living chemical compound into a living organism has been studied and investigated by Kepa  et al., (2017)\cite{7k}. As Cooper  (2019)\cite{12c} notes, simple organic molecules could form and spontaneously polymerize into macromolecules under the conditions thought to exist in primitive Earth's atmosphere. At the time life arose, the atmosphere of Earth is thought to have contained little or no free oxygen, consisting mainly of $CO_2$ and $N_2$ along with smaller amounts of gases such as $H_2$, $H_2S$, and $CO$. Such an atmosphere provided reducing conditions in which organic molecules, given a source of energy such as sunlight or electrical discharge, could form spontaneously. Kolb (2015)\cite{16k} also reviewed important turning points in chemical evolution that led to life, highlighting the complexity of this process and the need for continued research and investigation.

The concept of self-sustaining protometabolic cycles and self multiplying protocellular compartments is a critical aspect of the origin of life. These processes are believed to have played a pivotal role in the emergence of the first living systems.
Self-multiplying protocellular compartments refer to structures that are capable of replicating themselves through simple physical processes. These compartments are thought to have provided a means of protecting and replicating the self-sustaining protometabolic cycles.
The integration of these two processes, self-sustaining protometabolic cycles and self-multiplying protocellular compartments, is thought to have led to the emergence of the first living systems. This integration allowed for the exchange of energy and materials between the protometabolic cycles and the protocellular compartments, leading to the development of more complex and efficient metabolic processes.
The precise details of how these processes occurred and how they led to the emergence of life are still the subject of much research and debate. However, the concept of self-sustaining protometabolic cycles and self-multiplying protocellular compartments provides a framework for understanding the early steps in the origin of life and the emergence of the first living systems, for more informations see    Monnard et al.,  (2015)\cite{20mo}  and   Nader et al., (2022)\cite{24n}.

For many years, philosophers and scientists have been investigating the origins of life and seeking to understand the fundamental differences between living and non-living beings. They seek to answer questions such as how a non-living chemical compound transforms into a living organism and what unique characteristics living organisms share that distinguish them from non-living matter.
These questions are particularly challenging because the line between living and non-living matter is not always clear-cut. However, through careful observation and analysis, researchers have identified several key features that are characteristic of living organisms. These include the ability to grow, reproduce, respond to stimuli, maintain homeostasis, and undergo metabolism.
Despite these characteristics, the exact nature of what makes something "alive" remains an area of ongoing investigation and debate. By studying the origins of life and the properties that distinguish living from non-living matter, researchers hope to gain a deeper understanding of the fundamental nature of life and its place in the universe.

To facilitate research into the chemical origins of life,
the defining features of  cells are often divided into
hallmarks, such as growth, division, information processing,
and compartmentalization.
However, exactly how
these fundamental features of life could have emerged in
mixtures of non-living molecules remains one of the biggest
mysteries in cognitive science and is an active area of research in
chemistry,   biology and philosophy.
 In this manuscript,  we show that life is a purely physicochemical phenomenon.
There are many adequate expertise and experimental conditions, it is possible to synthesize almost any organic molecule in the laboratory under simulated prebiotic conditions which in the following, we will mention some of these sources.

Baum (2018)\cite{1bb} explores a generic description of life and uses it to explain how the chemically unusual life forms we observe today could have emerged. One of the most remarkable features of life is its apparent ability to defy the second law of thermodynamics and become increasingly out of equilibrium with its environment over time. To understand this phenomenon, Baum proposes a conceptual framework in which tendencies can be analyzed in relation to the expected equilibration point of the environment. He suggests that living systems correspond to metastable attractor states that tend to remain out of equilibrium with their surroundings.
Baum applies this framework to investigate the emergence of life from non-life, its evolutionary trajectory towards complexity, and the causes and impacts of cellular encapsulation. By viewing life as a metastable attractor state, Baum provides a novel perspective on how living organisms can persist in a state of disequilibrium with their environment, and how this disequilibrium may have contributed to the development of complex life forms.

The origins of life research has long grappled with a fundamental question: how were the proteinaceous side chains and the protein backbone selected during the early stages of evolution?  Frenkel-Pinter et al. (2019)\cite{13f} conducted a study on the oligomerization reactions of a group of positively charged amino acids, both proteinaceous and nonproteinaceous. Interestingly, these amino acids were found to spontaneously oligomerize without the need for enzymes or activating agents, under mild hydroxy acid-catalyzed dry-down conditions.
The findings presented in this paper are crucial for experimentally evaluating models of early protein evolution and enhancing our understanding of the origins of life. Models proposing chemical evolution in which unactivated amino acids directly condense to form polypeptides face several significant challenges, some of which have been partially addressed.

Protocells are cell-like compartments that mimic certain characteristics of living cells and are considered potential precursors to modern life. According to Cornell et al. (2019) \cite{12cc}, the initial protocells on early Earth likely developed with self-assembled membranes composed of fatty acids. However, a significant obstacle in comprehending the origin and resilience of protocells lies in the instability of fatty acid membranes when exposed to high salt concentrations or divalent cations, which would have been abundant in the early oceans.

During the process of converting a non-living chemical compound into a living organism, there is an intermediate state known as the "pseudo-alive" state. This state is characterized by the sensitivity of the chemical composition to the environment, which gives rise to properties that are characteristic of both living and non-living matter.
In the pseudo-alive state, the chemical compound exhibits several characteristics that are typical of living organisms, such as the ability to undergo self-replication and to respond to external stimuli. However, it also retains certain features of non-living matter, such as a lack of metabolic processes.
The pseudo-alive state represents a transitional phase in the evolution of life, in which non-living matter begins to exhibit properties that are characteristic of living organisms. By studying this intermediate state, researchers can gain insights into the fundamental nature of life and its emergence from non-living matter.

Pross, in his (2021)\cite{23pp} work, proposes that a cognizant chemical system that was capable of evolving and adapting to better exploit its environment would discover new possibilities for structural and organizational complexity, leading to the emergence of increasingly persistent forms. This process ultimately culminated in the emergence of the bacterial cell, which represented a significant milestone in the evolution towards greater persistence. Over the course of several billion years, this process led to the emergence of animals with neural systems and, ultimately, to the discovery of mind. Throughout this evolutionary process, the drive towards increasing persistence governed the development of life.
Pross's perspective offers a unique way of thinking about the origins and evolution of life on Earth, emphasizing the role of persistence in driving the emergence of increasingly complex organisms. By viewing life as a conscious chemical system, Pross highlights the importance of adaptation and evolution in the development of life on our planet. Overall, Pross's work provides valuable insights into the fundamental nature of life and its relationship with the environment.

The process of chemical sensitivity to the environment leads to molecular changes within the pseudo-alive state, resulting in increased sensitivity. Continued progression of this process ultimately prepares the system of chemical synthesis to become a living organism, creating unique characteristics that we experience in our consciousness as the special property of a living being. This experience gives rise to a fundamental distinction between living and non-living beings.
Notably, evolution occurs in all three stages of non-living, pseudo-alive, and living, although it is difficult to define clear boundaries between these stages. In Section 3, mathematical methods are used to demonstrate that there is no strong distinction between living and non-living beings.
This perspective highlights the role of chemical sensitivity in the evolution of life and emphasizes the continuity between non-living, pseudo-alive, and living states. 

In this article, the term "object" refers to the entity itself and its inherent characteristics, rather than the particles or organisms associated with it. For instance, an automaton is not considered a living being, even if it has bacteria, germs, or viruses on or within it. Similarly, a piece of meat  is not considered a living being based on its own characteristics, even though it may contain living organisms with the attributes of living beings. The focus here is on the essential qualities and properties of the object itself, independent of any external or internal components.

The mathematical concepts  used in this article are very simple and at the level of basic mathematical knowledge.

\section{The quest for the definition of the living being }  
When discussing living organisms, we typically think of beings that are similar to animals or plants. In order to define what constitutes a living organism, we must identify those characteristics that are shared by both plants and animals and that are considered essential from our perspective.
One hypothesis suggests that living organisms are characterized by the ability to undergo self-replication, respond to stimuli, maintain homeostasis, and undergo metabolism. Others have proposed that life is defined by the presence of certain biomolecules, such as DNA or RNA, or by the ability to evolve through natural selection.
Defining the distinction between living and non-living matter is a complex and ongoing task, and there are multiple competing hypotheses regarding the definition of life and the origins of life. Many researchers have proposed different frameworks for understanding the nature of life, and some of the most prominent hypotheses are outlined below.

Bartlett et al.,   (2020)\cite{1bd} seek to reframe the definition of life in a more expansive way while recognizing the need
to signify the specific kind of life that earthly forms represent. Thus, they have come up with a new
term-lyfe. Henceforth, they will refer to life (as we know it) and lyfe (as it could be, in the most general
sense). The two designations are distinguished as follows:\\
1)  Life represents life as we know it; it uses the specific disequilibria and classes of components of
earthly life. Life is an autocatalytic network of organometallic chemicals in aqueous solution
that records and processes information about its environment in molecular form and achieves
dynamical order by dissipating any subset of the following disequilibria: redox gradients,
chemiosmotic gradients, visible/thermal photons, etc.\\
2)  Lyfe represents any hypothetical phenomenon in the universe that fulfills the fundamental
processes of the living state, regardless of the disequilibria or components that it harnesses or
uses. Lyfe is any hypothetical phenomenon that maintains a low-entropy state via dissipation
and disequilibria conversions, utilizes autocatalytic networks to achieve nonlinear growth
and proliferation, employs homeostatic regulatory mechanisms to maintain stability and
mitigate external perturbations, and acquires and processes functional information about
its environment.

Gaining a precise understanding of the distinction between living organisms and non-living matter necessitates a comprehensive examination of the evolution of living beings and their origins. According to Kamila et al. (2020)\cite{20mv}, the initial phase of life's emergence involved a primitive nonenzymatic form of metabolism catalyzed by naturally occurring minerals and metal ions. This perspective, known as the "metabolism first" hypothesis, suggests that a primitive version of metabolism capable of synthesizing and breaking down ketoacids, sugars, amino acids, and ribonucleotides would be required for continuity with modern metabolism.
In their review, Kamila et al. provide accessible insights for chemists into the metabolic pathways relevant to the origin of life. The review also highlights experiments that propose several pathways may have originated from prebiotic chemistry. By exploring these pathways, we can gain valuable knowledge about the chemical processes that may have contributed to the emergence of life.

According to Pascal et al. (2013)\cite{23paa}, any scientific explanation of the origin of life must account for a driving force that can clarify how intermediate forms can remain stable over extended periods of time in a state that is far from equilibrium. In order to solve the question of life's origin, it is necessary to explain how states, which are considered unstable from a statistical and thermodynamic perspective, can acquire an alternative form of stability that enables further improbable changes. Pascal et al. introduced a concept called dynamic kinetic stability, which is specific to entities capable of self-reproduction. This new form of stability provides a satisfactory explanation for the processes governing transformations in both inanimate and animate systems. Each form of stability is supported by its own unique mathematical logic.

One form of stability is thermodynamic stability, which dominates the regular chemical world and has been understood since Boltzmann. It involves the tendency of physico-chemical systems to move towards more probable states. In contrast, dynamic kinetic stability is a distinct form of stability that is specific to persistent replicating systems. It arises from the dynamic persistence associated with exponentially driven self-replication. Understanding the interrelation between these two distinct forms of stability, each with its own mathematical logic, is crucial for explaining the essence of biology.

According to the study conducted by Pavlinova et al. (2022)\cite{23po}, the transition from non-living matter to living matter likely required some form of evolutionary process. The complexity of molecular structures necessary for a chemical system to function at a level comparable to life poses a significant challenge for their spontaneous emergence. Additionally, the vastness of chemical possibilities further complicates the likelihood of such a development. Consequently, researchers have explored potential intermediate stages and processes that could facilitate a gradual transition, with a particular focus on systems capable of Darwinian evolution.
In their paper, Pavlinova  presented an alternative hypothesis suggesting that catalytically facilitated recombination might have played a crucial role in enabling a more rudimentary form of Darwinian evolution. This mechanism could have provided the means for the emergence and development of simpler life forms.

Ameta et al. (2021)\cite{1aaa} discovered that network-level structures play a critical role in constraining the properties that emerge at the level of chemical compositions. The researchers found that network topology imposes trade-offs between growth, which is favored by higher connectivity, and variation, which is favored by lower connectivity. Similarly, there are trade-offs between variation, which is favored by lower connectivity, and robustness to environmental changes, which is favored by higher connectivity.
These findings have direct implications for scenarios in which early evolution is driven by environmental heterogeneity. The trade-offs imposed by network topology are likely to have played a significant role in shaping the evolution of early life forms, determining which organisms were better suited to their environment and which were not. By understanding the relationship between network-level structures and the emergence of properties at the chemical level, researchers can gain valuable insights into the fundamental nature of life and its origins.
As described in the Ameta et al. (2021) paper, in prebiotic scenarios such as Dynamical Kinetic Stability, the process of evolution depends on a delicate balance between the persistence of chemical compositions (i.e., robustness to environmental changes) and the exploration of novel compositions (i.e., susceptibility to perturbations). These evolutionary trade-offs, along with the connectivity rules imposed by network topology, suggest the existence of a "Goldilocks" range, in which the density of catalytic interactions is neither too high nor too low, for evolution to occur.
The researchers' findings suggest that the emergence of life is not solely a matter of chance, but is instead subject to fundamental constraints imposed by the underlying network-level structures. By investigating the interplay between network topology and the emergence of properties at the chemical level, researchers can gain a deeper understanding of the mechanisms that drive the evolution of life and the origins of living systems. Ultimately, this knowledge can shed light on the complex interplay between matter and consciousness that underlies the nature of existence.

Researchers can investigate the interplay between network topology and the emergence of properties at the chemical level through a combination of theoretical modeling and experimental studies, for example see Winterbach et al. (2013)\cite{33ww} and Somarakis et al. (2016)\cite{8ccc}.
Theoretical models can be used to simulate the behavior of complex networks and their impact on the emergence of properties at the chemical level. These models can take into account various factors such as the connectivity of the network, the chemical reactions that occur within the network, and the environmental conditions that affect the network's behavior. By using theoretical models, researchers can explore the impact of different network topologies on the emergence of properties at the chemical level and gain insights into the constraints and trade-offs that shape the evolution of living systems.
Experimental studies can also provide valuable insights into the interplay between network topology and the emergence of properties at the chemical level. Researchers can design experiments to manipulate the connectivity of chemical networks and observe the resulting impact on the emergence of properties such as growth, variation, and resilience to environmental changes. By conducting these experiments, researchers can test the predictions of theoretical models and gain a more comprehensive understanding of the fundamental mechanisms that underlie the origins and evolution of life.
In summary, investigating the interplay between network topology and the emergence of properties at the chemical level requires a combination of theoretical modeling and experimental studies. By using these complementary approaches, researchers can gain deeper insights into the complex interplay between matter and consciousness that underlies the nature of existence.

The dynamics observed in the system under investigation highlight the non-obvious role of self-assembly in driving self reproduction. While self-assembly has been previously demonstrated to drive self reproduction in other systems, such as tubular assemblies of molecules, its role in the system under investigation was not immediately apparent.
The robustness of self-assembly as a mechanism for driving self-reproduction is due to its ability to create organized structures from simple building blocks. This process can occur spontaneously, without the need for external energy input. The resulting structures can then serve as templates for the formation of new structures, leading to self-replication and self-assembly.
In the system being studied, the role of self-assembly in driving self-reproduction was not initially clear. However, the observed dynamics suggest that self-assembly may play a critical role in driving the emergence of self-replicating structures. By understanding the underlying mechanisms that drive self-assembly and self-reproduction, researchers can gain valuable insights into the origins and evolution of life, for more informations see  Pross  and  Pascal (2017)\cite{23p};  Merindol  and  Walther (2017)\cite{20mv} and Muchowska et al. (2022)\cite{20mv}.

According to Belthle and Tüysüz (2022) \cite{1bt}, under-sea hydrothermal vents offer a plausible geochemical environment that could have facilitated the formation of reduced carbon compounds during prebiotic metabolism at the emergence of life. These vents, which produce hydrogen gas (H2), create a suitable setting for the prebiotic synthesis of the first organic compounds.

In line with the "metabolism-first" hypotheses regarding the origin of life, the initial stages involved the emergence and evolution of proto-metabolisms, which are organizations of molecular species. Unlike the involvement of self-replicating RNA, these proto-metabolisms are believed to have emerged independently. They are characterized as self-producing and self-amplifying because the formation of each member of the metabolism is catalyzed by at least one other member within the metabolism. Additionally, they have access to a suitable source of free energy and other necessary resources. As a collective network of molecular species, they exhibit autocatalytic properties.

The constituents of the proto-metabolism work together to amplify the formation of each other. This metabolic organization can include catalytic polymers, small-molecule autocatalytic cycles, and processes. Under specific conditions, these components cooperate to facilitate the emergence and evolution of the proto-metabolism. For more detailed information on this topic, refer to the manuscript by Stewart (2018)\cite{24st}.

The search for a precise definition of living beings has been a topic of study and investigation for philosophers and scientists for a long time.
Methodological questions also arise over the most appropriate approach to
elaborating and justifying any definition of life (Cleland and Chyba 2007\cite{10a}).
  As mentioned by Malaterre and  Chartier (2021)\cite{20mc}, 
  should “life” be defined on the basis of the common properties shared by
particular instances of life? But in this case, how should such instances of life be
identified in the first place? Or should “life” be first defined, with the resulting
definition only then being used to identify particular instances of life? Yet in that
case, where would such a definition come from and how would it be justified?
The “problem of defining life” can be explicated as an ontological
problem. In that case, questions of interest typically concern whether life picks
out a definite natural kind by delineating entities that are alive in some sense from
entities that are not, or whether some specific account of natural kinds is better
suited than another to capture “life”.

Many philosophers various aspects of the definition of living organisms have been investigated.
   Cleland (2012, 2019)\cite{10c, 11c}   presents a comprehensive and nuanced argument in favor of pursuing universal biology despite the complex and diverse nature of life on Earth, which William James vividly described as a "blooming buzzing confusion". In her work, Cleland advocates for the exploration of universal biology, which involves studying the possibility of life forms and processes that may exist beyond Earth. She acknowledges that the study of life on Earth can be intricate and bewildering, with its vast array of species, ecosystems, and biological phenomena. However, instead of viewing this complexity as a deterrent, Cleland sees it as a motivation to expand our understanding of life in a broader context.
Cleland argues that by solely focusing on Earth-based life, we may limit our understanding of the potential diversity and complexity of life forms that could exist elsewhere in the universe. She emphasizes the importance of interdisciplinary approaches and collaboration between scientists from various fields, including biology, chemistry, physics, and philosophy.

                     Knuuttila  and   Loettgers  (2017)\cite{14k}    have  studied on the comprehensive definition of living being, but they did not reach a general conclusion to separate living beings from non-living. 
  Malaterre  (2010)\cite{20ma}   proposed two arguments: first, that the roots of the tree of life extend well beyond the commonly recognized "ancestral organisms" and include much simpler, minimally living entities that can be referred to as "protoliving systems"; and second, that these roots gradually dissipate into non-living matter along several functional dimensions. Between non-living and living matter, there exist physico-chemical systems that exhibit a "lifeness signature." This signature could also explain a variety of biochemical entities that are considered to be "less-than-living" but "more-than-non-living.

Based on the sources we have mentioned, there are two main reasons why it can be difficult to distinguish between living and non-living things. The first reason is that for some organisms, the characteristics of life and non-life can occur at different times and under different conditions. In other words, these organisms can exist in a state of dormancy or suspended animation where they do not exhibit the typical signs of life, such as metabolism or reproduction. However, under the right conditions, they can resume their living state. This makes it challenging to classify them as strictly living or non-living.
The second reason is that there is no single defining characteristic that sets living things apart from non-living things. Instead, living organisms exhibit a combination of characteristics such as organization, metabolism, growth, reproduction, adaptation, and response to stimuli. While non-living things may exhibit some of these characteristics, they do not exhibit all of them in a coordinated and integrated way. This makes it difficult to draw a clear line between living and non-living entities.
On the other hand,  we categorize certain creatures as living beings and the rest as non-living, but this distinction is arbitrary and without meaning. By separating a group of beings based on certain characteristics, we imply that the remaining beings lack these characteristics, but this assumption is flawed. Without a precise definition of the characteristics that distinguish living beings, and without a comprehensive understanding of other beings that may or may not possess these characteristics, our categorization is incomplete and potentially inaccurate.

From the point of view  Jianhui (2019)\cite{7j}, 
The definition of life may conflict with, or even contradict, our common-sense understanding of the term. Our common-sense concept of life is typically associated with the general characteristics of animals and plants, such as growth, reproduction, self-sustainment, and responsiveness to external stimuli. However, if we wish to provide a comprehensive definition of organisms, we must consider the characteristics of all types of life, including microorganisms like bacteria, as well as viruses, viroids, and prions.
 The characteristics of these creatures are very different from those of more commonly conceived organisms. 
Physiology, for example, often defines organisms systems that perform functions, such as digestion, metabolism, excretion, respiration, movement, growth, development, and response to external stimuli as living systems.
 Biochemistry and molecular biology often regard living organisms as systems that can transmit genetic information encoded in  DNA  and Ribonucleic Acid (RNA), which can control the synthesis of proteins, which determine the main properties of organisms. 
 Of course some organisms, such as viruses, are between living and nonliving in many ways.

As previously mentioned, the concepts of living beings and life phenomena are forms of conscious experience.
At a certain stage of development, newborn babies begin to differentiate between creatures that can move or react to their surroundings and those that cannot. The separation of living and non-living beings based on behavior is a gradual process that is carried out by humans.
Ultimately, the human conscious experience creates a distinction between living and non-living beings. However, the question arises as to whether this distinction has a logical foundation or if it is merely an illusion. Can we define a characteristic that completely separates living beings from non-living matter?
In the next section, we will explore the possibility of identifying a complete sample of all living organisms that share common characteristics that satisfy our criteria.

 \section{Mathematical method for examining the distinction between living and non-living beings}
 The interaction of non-living chemical compounds, $\gamma$,  with the environment, a cognitive system in $\gamma$, which possesses the ability to process information and adapt, has evolved. This cognitive system gradually became more complex over time, leading to the emergence of living organisms with intricate biological structures and functions. As a result of the evolution of this cognitive system, living organisms have been created. There are compelling reasons behind this phenomenon, some of which have been mentioned previously, while others are outlined below.

According to Pascal and Pross (2022)\cite{23pa}, all material systems, whether living or non-living, exhibit responses to external perturbations, albeit in distinct ways. Non-living matter, for instance, responds to perturbations through the directing effects of the Second Law of Thermodynamics. For example, heating a physical object can cause it to expand or undergo a phase transition. Similarly, perturbing a chemical system at equilibrium, such as introducing an additional reagent or altering the concentration of its constituents, leads to the system seeking to re-establish its equilibrium state. In both physical and chemical cases, the system's response to the perturbation can be explained using thermodynamic principles.
In contrast, living systems demonstrate different behaviors. While they adhere to the physical laws of nature, they primarily operate according to biological principles and respond to perturbations through a phenomenon known as adaptation. Organisms adjust to perturbations in a manner that aligns with their own agendas or survival strategies.
Pascal and Pross also discussed how the replicative Dissipative Kinetic Selection (DKS) state, which is common to both chemical and biological systems, could shed light on the physical basis of cognition and the potential origins of mental states in living organisms.

Pascal and Pross's exploration of the replicative DKS state provides insights into the physical basis of cognition and the potential origins of mental states in living organisms. The replicative DKS state refers to a state in which a system undergoes self-replication while simultaneously dissipating energy. This concept is applicable to both chemical and biological systems.
In the realm of chemistry, the replicative DKS state can be observed in autocatalytic reactions. These reactions involve the generation of a catalyst that accelerates its own production, leading to self-replication. The dissipative aspect of the DKS state is associated with the energy dissipation that occurs during the reaction. 

By establishing a connection between the physical principles underlying chemical self-replication and the behaviors observed in living organisms, Pascal and Pross propose that cognition and mental states may have originated from the fundamental processes of self-replication and energy dissipation. This perspective implies that the ability of living systems to adapt and respond to perturbations is rooted in the same principles that govern self-replication in chemical systems.

On the other hand,  drawing on Shapiro works (2007 and 2021)\cite{24shap1, 24shap}, it is posited that every living being possesses a cognitive system that gives rise to a unique form of consciousness. Primitive organisms, under specific conditions or chemical states, can develop such cognitive systems. From this perspective, the chemical evolution of life can be seen as a complex kinetic process that engenders a distinct state within our consciousness system. This state allows us to map the emergence and evolution of life as a cognitive system, which can be understood as a thermodynamic phenomenon. Thus, a comprehensive understanding of the functioning of cognitive systems in various organisms can enrich our comprehension of the human consciousness system.
Shapiro argues that all living organisms, including plants, possess cognitive systems. However, the prevailing assumption is that plants lack consciousness, primarily due to a top-down perspective that employs human consciousness as the benchmark for assessing consciousness in other beings. This approach fails to consider the fundamental differences between plants and animals, resulting in the rejection of plant consciousness based on criteria derived from animal consciousness. 
Shapiro's argument is rooted in the belief that cognition extends beyond the boundaries of the central nervous system and can be found in a wide range of organisms. He proposes that cognition should be understood as a distributed phenomenon, where various parts of an organism interact and process information to guide adaptive behavior. In the case of plants, this distributed cognition manifests through complex physiological processes, such as signal transduction, environmental sensing, and response mechanisms.

The emergence of living organisms' systems can be attributed to internal chemical and physical changes in their composition,  $\gamma$, in response to environmental factors. The $\gamma$ cognitive system has evolved not only in living biological states but also in non-living chemical states. Put simply, the evolution of cognitive systems in living organisms begins from a non-living chemical state and progresses into a living biological state. In this chapter, we demonstrate, using mathematical methods, that there is no clear boundary separating these two states.
Certainly, from my perspective, the biological evolution of living organisms can be regarded as a distinctive manifestation of chemical and physical evolution. In essence, the creation and subsequent evolution of living organisms encompass a broader framework of chemical and physical processes.
Thus the process of biological evolution involves the gradual change and diversification of living organisms over time. It is driven by various mechanisms such as genetic mutation, natural selection, genetic drift, and gene flow. These mechanisms operate within the framework of chemical and physical laws that govern the behavior of matter and energy.
At its core, life itself is fundamentally based on chemistry. The intricate molecular interactions and biochemical processes that occur within living organisms are governed by the principles of chemistry and physics. From the formation of complex organic molecules to the functioning of cells, tissues, and organ systems, the underlying mechanisms can be explained by the laws of chemistry and the physical properties of matter.

In their insightful work, Kolb (2016)\cite{16k} delved into key subjects that are crucial for comprehending the chemical origins of life. The author also emphasized the role of philosophical methodologies in defining life and gaining a deeper understanding of the transition from abiotic to biotic states.
One notable milestone discussed by Kolb is Oparin's model, which provides valuable insights into the initial stages of chemical evolution predating the existence of RNA-based systems. Oparin's model proposes the spontaneous formation of coacervates, which encapsulate chemical matter, exhibit primitive self-replication, and pave the way for rudimentary metabolism. In addition to discussing Oparin's model, the author also reviews recent experimental breakthroughs in our laboratory that build upon this model. Furthermore, she explores the potential types of selection that may have operated within these early systems.
Another significant milestone in chemical evolution is the transition from abiotic to biotic systems. This transition occurred after the development of the RNA world, marking a crucial shift in the emergence of life.

 According to Pascal and Pross (2016)\cite{23paaa}, 
  life as a  phenomenon can be seen to rest on an extrathermodynamic
(kinetic) base, one that derives directly from mathematical/logical considerations.
Life, first and foremost, is a self-sustained replicative network of chemical reactions whose
evolutionary roots lie in some simple primordial replicative system whose identity has long faded
in the mists of time. But once such a simple (but persistent) replicating system was able to emerge
from the materially diverse environment that was manifest on our planet some four billion years
ago, the logic of the persistence principle, supported by the math of exponential growth, led
inevitably to an evolutionary process of increasing complexity, both within the individual
replicating forms (protocells and cells), and through network formation between those individual
forms.

 Pross  and  Pascal (2017)\cite{23p}   attempted to explain life and its emergence within a
general physicochemical context. Once it is appreciated that life
emerged from inanimate beginnings in a well-defined process
with an identifiable driving force. The biological and
physical worlds are intimately connected through process.
There is a process, explicit and physicochemically defined, that
under appropriate contingent conditions, leads from chemistry
to biology such that these two worlds merge into one. So,
though life is a complex chemical system exhibiting complex
kinetic behaviour, that complex behaviour can be traced back to
self-reproducing chemical systems maintained far from equilibrium
and directed by kinetic driving forces.  This paper attempts to place life and its emergence within a
general physicochemical context. Once it is appreciated that life
emerged from inanimate beginnings in a well-defined process
with an identifiable driving force. The biological and
physical worlds are intimately connected through process.
There is a process, explicit and physicochemically defined, that
under appropriate contingent conditions, leads from chemistry
to biology such that these two worlds merge into one.

 Pascal and Pross (2022)\cite{23pa} introduced an intriguing concept of a non-equilibrium state of matter called an energized dynamic kinetic state, which challenges the prevailing understanding of the emergence of life from non-life. They demonstrated that certain chemical systems, when activated into this dynamic kinetic state, can exhibit rudimentary cognitive behavior. This proposition challenges the conventional notion that life emerged solely from non-living matter.
The exploration of the energized dynamic kinetic state by Pascal and Pross suggests the existence of alternative pathways for the emergence of cognitive behavior. It challenges the traditional view that mental capabilities are exclusive to living organisms. By activating specific chemical systems into this state, they propose that basic cognitive processes could manifest, blurring the line between the living and non-living.
This research carries profound implications for our understanding of the origins of life. It urges us to reconsider the idea that life arose solely from non-living matter through a gradual accumulation of complex chemical reactions. Instead, Pascal and Pross propose that cognitive phenomena might have been present even in the early stages of chemical evolution.

Regarding the transition from a non-living to a living chemical state, extensive research has been conducted, as an examples see
 Merindol and  Walther (2017)\cite{20mm}; 
Lehn	  (2007)\cite{19ll};
Ziemann et al. (2009)\cite{33};
Liang  (2016)\cite{19lll}; 
Agozzino  et al. (2020)\cite{1a};
Kristin et al. (2004)\cite{16kk}; 
  Greer et al. (2016)\cite{13G};    Krämer  et al. (2022)\cite{8cc};   Kaklauskas et al. (2022)\cite{7jj};   Jeziorski  et al. (2022)\cite{7};   Slootbeek et al.  (2022)\cite{24ss};    Stewart (2019)\cite{24st},   Trefil et al. (2009) \cite{31t},  Jeancolas, et al. (2020)\cite{Jeancolas}, and Pascal and Pross (2022)\cite{23pa}.
 
In the following, by mathematical methods, I  show that there is no comprehensive definition of living being that completely distinguishes them from non-living being. 
 
Suppose that we divide the creatures of the universe into two distinct groups living and non-living creatures which are not similar to each other at a point in time compared to the characteristics we defined for living organisms and call them $A$ and $B$, respectively. The basis for this distinction into two distinct groups is to define properties for members of  $A$. 
Now if our assumption is that we have correctly identified and defined these characters, for this separation we need to make sure that  members of $B$ do not have all these characters, and for this case, it is necessary that we recognize all members of $B$ that is not possible.
On the other hand, elements $A$ can easily leave group $A$ and enter group $B$ over time, and vice versa.
So, this problem makes  difficult to  comprehensive definition for living beings.
But despite these problems, we do not stop here, and we can take living beings as an example of  systems that includes humans, animals, plants, bacteria, and   maybe viruses.

In the definition of living organisms, we can consider two cases that all living being have at least one common characteristic or each category of living being has a common characteristic.
If we want to study the problem in mathematical language, suppose we have a finite number of properties for all living organisms that we denote by $$q_1(x), q_2(x), q_3(x),\cdots,q_n(x), $$ where some of the properties $q_1, q_2, q_3, \cdots,q_n$ holds for a living being $x$. On the other hands, if some of the properties   $q_1, q_2, q_3, \cdots,q_n$ holds for $x$, then $x$ is living creature or $x\in A$. Put
\begin{equation}
A=\{x\in U: ~~q_i(x) ~\text{holds} ~\exists i(1\leq i\leq n)\}.
\end{equation}
As I said that $A$ is including  all of living creature. 
The properties of $q_i$   determine members of $A$, or, conversely, the presence of members $A$  determine  property $q_i$ for some $(1\leq i\leq n)$.
There is an another problem in the above definition that   a number of organisms which identified as non-living may be members of set $A$.
For example,  dose artificial intelligence machines, viruses or amino acids may belong to $A$?

In this case, we define the set $A$ that  all members have at least one common characteristic.
In order to define living organisms, we need their complete samples of living organisms. In this regard, we only have a number of samples with similar characteristics which including humans, animals, plants,   microorganisms and others.
Let we put the samples  of all living organisms into  categories  $A_1$, $A_2$, ..., $A_k$ where we can assume that $A_1$ be a group of mammals, $A_2$ be a group of insects, $A_3$ be a group of plants and others  which this sample of living being is complete, that is, $A=\cup_{i=1}^k A_i$.

 Let $\{q_1, q_2, q_3,\cdots,q_n\}$ be a finite sequence of special characterization of all organisms members of $A$.   
 Let $\Omega\subseteq \{q_1, q_2, q_3,\cdots,q_n\}$ be non-empty such that all living beings satisfies to all characteristics of the elements of the set $\Omega$. Assume that $\Omega=\{q_{k_1}, q_{k_2}, q_{k_3},\cdots, q_{k_t}\}$ where $1\leq t \leq n$. 
 Our main that  $x$ satisfies  in the properties $\Omega$ is equivalent to, $x$ satisfies in the properties $q_{k_1}\wedge q_{k_2}\wedge q_{k_3}\cdots \wedge q_{k_t}$,  equivalently $q_{k_1}(x)$,  $q_{k_2}(x)$,  $q_{k_3}(x)$,...,  and $q_{k_t}(x)$ hold.
 Suppose $U$ is the collection of elements 
 on the world, then

 $$A=\{x\in U:~x~\text{satisfies ~in ~the~properties}~\Omega\}.$$  
 Now let $b_{(k_1, k_2,\cdots k_t)}= q_{k_1}\wedge q_{k_2}\wedge q_{k_3}\cdots \wedge q_{k_t}$. So $b_{(k_1, k_2,\cdots k_t)}(x)$
  holds means that $x$ is a living being or equivalently $x\in A$.   
It is clear that we can display the set $A$ as below
  $$A=\{x\in U: ~~b_{(k_1, k_2,\cdots k_t)}(x)~\text{holds} ~\}.$$
For the existence of the set $A$, it must be $\Omega\neq \emptyset$. 
The characteristics that we mentioned above for living organisms are exactly dependent on the  samples of living organisms, but these samples, especially the complete samples of living organisms are selected based on their characteristics. As a result, the characteristics of living organisms and the selection of a complete sample cannot be independent of each other. So $A$ and  $\Omega$  completely dependent on each others.

Based on the information provided from various sources, it is presumed that a chemical compound, denoted as $\gamma$, (without any ambiguity and losing generality, we assume that these compounds  is a zygote such as egg, plant seeds etc)   does not possess the characteristics associated with living organisms (and is therefore not a member of the set $A$). However, it is postulated that through chemical and physical processes occurring in the environment surrounding $\gamma$ (such as incubation time or other processes), it undergoes a transformation into a living entity (thus becoming a member of the set $A$). We will denote the physical, chemical, and biological characteristics of $\gamma$ as $a$, and the corresponding characteristics of the transformed entity as $b$.

Now we define a function $f$ from $[a,b]$ onto $[0,1]$ where $f(a)=0$,  $f(b)=1$ and $f(x)$ is a number  which shows the distance between simple $\gamma$s and a stage of embryo formation for $a\leq x\leq b$. More precisely, if we show the start time of $\gamma$ incubation
  with $t_a$ and  also represent the moment when the chick hatches with $t_b$, then the position $a$ occurs at time $t_a$,  the position $x$ occurs at time $t_x$ and  the position $b$ occurs at time $t_b$.
So, we define $f(x)=d(a,x)=\frac{\vert t_x-t_a\vert}{\vert t_b-t_a\vert}$  where $d$ is a distance from $a$ to $x$ for all $x\in [a,b]$.
The function $f$ is continuous, and so we cannot determine when the number $f(x)$ must be in order for $x$ to be living in our definition.
If we define the set $$\omega=\{f(x):~ x~\text{is}~ \text{a} ~\text{living} ~\text{organism}\}.$$
Then $1\in \omega$.  So $\omega$ is non-empty and bounded below. Let $\inf \omega=\alpha$. Then there is $x_0\in [a,b]$ such that $f(x_0)=\alpha$. It follows that for all $x,y\in [a,b]$ with $f(x)<\alpha<f(y)$, the position of $x$ indicates that the $\gamma$ is not alive (or $b_{(k_1, k_2, ...k_t)}(x)$ not holds) and the position of $y$ indicates that it is alive (or $b_{(k_1, k_2, ...k_t)}(y)$ holds). 
Consider the following diagram:
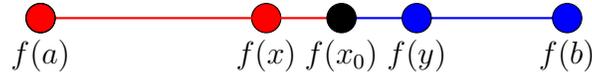
\begin{center}
\begin{tikzpicture}
\node[circle,draw,fill=red] (1) at (0,0) (3pt){};
\node[circle,draw,fill=red] (3) at (3,0) (3pt){};
\draw[thick,color=red] (0.19,0) -- (4,0);
\draw[thick,color=blue] (4,0) -- (7,0);
\node[circle,draw] (1) at (0,0) (3pt){};
\node[circle,draw,fill=black] (3) at (4,0) (3pt){};
\node[circle,draw,fill=blue] (7) at (5,0) (3pt){};
\node[circle,draw,fill=blue] (5) at (7,0) (3pt){};
\node at (0,-0.5) {$f(a)$}{};
\node at (3,-0.5) {$f(x)$}{};
\node at (4,-0.5) {$f(x_0)$}{};
\node at (5,-0.5) {$f(y)$}{};
\node at (7,-0.5) {$f(b)$}{};
\end{tikzpicture}
\end{center}

We can take $f(x)$ and $f(y)$ (or $t_x$ and $t_y$) so close together that the contradiction of $x$ and $y$ do not differ (regarding as the common characterizations mentioned in $\Omega$), so that  $\Omega=\emptyset$.

On the other hand, in the next scenario, if an $\gamma$ (zygote) has the characteristics of living organisms (i.e., it is a member of set $A$) but incubation is not performed, it will eventually become a non-living entity. 
Now, let us consider the time when the $\gamma$ is removed from set $A$. In this case, if we repeat the above argument, we can conclude that $\Omega$ is empty.
According to the explanations provided in the previous chapter, the process of converting a non-living chemical compound into a living organism or vice versa also follows this argument. However, if the members of set $A$ do not necessarily have a common characteristic, then we will have definition (3,1). 
However, the above arguments can lead to a contradiction, which highlights the difficulty of precisely defining the boundary between living and non-living entities. As a result, it becomes challenging to distinguish when a non-living chemical compound becomes a living organism.
For instance, consider the process of water changing from a liquid state to solid ice. We cannot specify a fixed time, such as $t_0$, at which the water transitions from being liquid to solid (ice). This example highlights the continuous and dynamic nature of the transition from non-living to living systems, which further complicates the process of defining the characteristics of living beings.

Charlat et al. (2021)\cite{8cc} argue for the bridging of the gap between physico-chemical and biological systems and explain the transition from inanimate to living matter. However, the material presented above suggests that there is no fundamental gap between inanimate and living matter in space and time.
The concept of pseudo-alive beings also plays an essential role in the connection between living and non-living properties. While there may not be a precise and complete definition of living beings that distinguishes them from non-living beings, the characteristics of pseudo-alive beings can help to bridge the gap between the two.
 According to the above argument, we can easily say that there is not and will not be a precise and complete definition of the creatures which called living beings that distinguishes them from non-living beings.

\section{Corollary}
The emergence of a cognitive chemical system that is able to evolve and adapt to better exploit its environment is a crucial step towards becoming a living being, as it enables the system to find new structural and organizational possibilities that increase sustainability.
However, the process of transitioning from non-living to living is not a clear-cut distinction, but rather a continuous process. This makes it difficult to precisely distinguish between the characteristics of non-living and living systems. 
Consequently, there is not and will not be a precise and complete definition of the creatures which called living beings that distinguishes them from non-living beings. That is, in a process of chemical transformation of a non-living being into a living one, we cannot have a specific and certain time before which this being is non-living and after that it is alive.
 Therefore, in the process of converting a non-living chemical variable into a living chemical entity, we cannot say exactly which fixed time this happens  or has already happened in this process. On the other hand, we do not have unique characteristics for living beings that completely distinguish them from non-livings.
So there is no real and logical wall between living and non-living beings.
 Pascal and Pross (2022)\cite{23pa} demonstrated that Darwin's theory of evolution extends beyond the realm of biology and encompasses non-living chemical systems as well. In this paper, I show  that the boundary between biological and non-biological states is not rigid, implying a continuous process of evolution across both domains.

\vspace{0.2cm}

{\bf Data availability.}  No data were used to support this study.

\vspace{0.2cm}
{\bf Conflicts of Interest.}
The authors declare that they have no conflicts of interest.


\begin{thebibliography}{1}


\bibitem{1a}    Agozzino, L.  et al.  (2020) How do cells adapt? Stories told in landscapes. Annu Rev Chem Biomol Eng.  11, 155-182. 
{\textcolor[rgb]{0.00,0.00,0.84}{https://doi: 10.1146/annurev-chembioeng-011720-103410}}  


\bibitem{1aa}     Allen, J.S.   (2009)   The lives of the brain
human evolution and the organ of mind, Harvard University Press.

\bibitem{1aaa}  
Ameta,  S.  et al. (2021) Darwinian properties and their trade-offs in autocatalytic RNA reaction networks. Nat Commun. 12, 842.
{\textcolor[rgb]{0.00,0.00,0.84}{ https://doi.org/10.1038/s41467-021-21000-1}} 





\bibitem{1bd}  
Bartlett, S. and  Wong, M.L.  (2020)  Defining lyfe in the universe: from three privileged functions to four pillars. Life.  16;10(4), 42. 
{\textcolor[rgb]{0.00,0.00,0.84}{ https://doi: 10.3390/life10040042. PMID: 32316364}} 




\bibitem{1bb}     Baum, D.A.  (2018) The origin and early evolution of life in chemical composition space. J Theor Biol. 456, 295-304. 
{\textcolor[rgb]{0.00,0.00,0.84}{https://doi: 10.1016/j.jtbi.2018.08.016}}  

\bibitem{1bt} 
 Belthle, K.S.  and  Tüysüz, H.    (2023)  Linking catalysis in biochemical and geochemical CO2 fixation at the emergence of life.  Chem. Cat. Chem.  15(4), 1-6.
{\textcolor[rgb]{0.00,0.00,0.84}{https://doi: 10.1002/cctc.202201462}}  








\bibitem{8cc}
Charlat, S. et al.  (2021) Natural Selection beyond Life? A Workshop Report. Life.  11(10), 1051.
{\textcolor[rgb]{0.00,0.00,0.84}{https://doi.org/10.3390/life11101051}}  
 


\bibitem{9c}   Clark, R.E. et al.  (2002) Classical conditioning, awareness, and brain systems. Trends
Cogn Sci.   6, 524–531. 
{\textcolor[rgb]{0.00,0.00,0.84}{https://doi.org/10.1016/ s1364- 6613(02) 02041-7}}  



\bibitem{10a} 
Cleland, C.E.  and  Chyba,  C.F. (2007)  Does 'life' have a definition? in Planets and Life: The Emerging Science of Astrobiology. Cambridge University Press. 119-131.

\bibitem{10c} Cleland,  C.E.  (2012)  Life without definitions. Synthese.  185(1), 125-144.  
{\textcolor[rgb]{0.00,0.00,0.84}{https://doi.org/10.1007/s11229-011-9879-7}}  




\bibitem{11c} Cleland, C.E.  (2019)  The quest for a universal theory of life, University Printing House.  Cambridge CB2 8BS, United Kingdom.  

 \bibitem{11cc}  Crick,  F.  and   Koch C (1990) Towards a neurobiological theory of consciousness.
Semin,  Neurosci.  2, 263–275.
{\textcolor[rgb]{0.00,0.00,0.84}{http://resource.nlm.nih.gov/101584582X469}}  

\bibitem{12c} Cooper,  G.M.  (2019)  The Cell: A molecular approach. Oxford University Press.






 \bibitem{12cc}     Cornell, C.E. et al.  (2019) Prebiotic amino acids bind to and stabilize prebiotic fatty acid membranes (2019) Proc Natl Acad Sci.  116(35), 17239-17244. 
 {\textcolor[rgb]{0.00,0.00,0.84}{http://doi: 10.1073/pnas.1900275116}}
 


  
\bibitem{13f}     Frenkel-Pintera,  M.  et al. (2019)  Selective incorporation of proteinaceous over non-
proteinaceous cationic amino acids in model prebiotic oligomerization reactions.
Proc Natl Acad Sci.  116(33), 16338–16346.
{\textcolor[rgb]{0.00,0.00,0.84}{https://doi.org/10.1073/pnas.1904849116}}  

\bibitem{13G}   Greer, P.L. et al. 
(2016) A family of non-GPCR chemosensors defines an alternative logic for mammalian olfaction. Cell. 165(7), 1734-1748.
{\textcolor[rgb]{0.00,0.00,0.84}{https:// doi: 10.1016/j.cell.2016.05.001}}  

 \bibitem{Jeancolas}
 Jeancolas, C. and  \ Malaterre, C.   Nghe, P. (2020) Thresholds in Origin of Life Scenarios, iScience. 23(11),   101756,
 {\textcolor[rgb]{0.00,0.00,0.84}{https://doi.org/10.1016/j.isci.2020.101756.   }}  
   
    
    
    
    
\bibitem{7}   Jeziorski, J.   (2022) Brain organoids, consciousness, ethics and moral status. Semin Cell Dev Biol.  144, 97-102. 
{\textcolor[rgb]{0.00,0.00,0.84}{https://doi: 10.1016/j.semcdb.2022.03.020}}  

      
\bibitem{7j} Jianhui,  L.  (2019)  On the definition of life.  Philosophy Study. 9(9), 497-511.  
{\textcolor[rgb]{0.00,0.00,0.84}{https://doi.org/10.17265/2159-5313/2019.09.001}}  



    
    

\bibitem{7jj} Kaklauskas,  A.  et al.   (2022) Review of AI cloud and edge sensors, Methods, and Applications for the Recognition of Emotional, Affective and Physiological States. Sensors.  22, 1-80.
{\textcolor[rgb]{0.00,0.00,0.84}{ https://doi.org/10.3390/s22207824}}  





\bibitem{7k}      Kepa, R-M.   et al. (2017) Chemical roots of biological evolution: the origins of life as a process of development of autonomous functional systems. Open Biol. Published by:Royal Society. 7(4), 1-10.
{\textcolor[rgb]{0.00,0.00,0.84}{https://doi.org/10.1098/rsob.170050}}  



\bibitem{14k} Knuuttila,  T. and   Loettgers,  A.  (2017)  What are definitions of life good for? Transdisciplinary and other definitions in astrobiology.  Biology  and  Philosophy.  32(6), 1185–1203. 
{\textcolor[rgb]{0.00,0.00,0.84}{https://doi.org/10.1007/s10539-017-9600-4}}  
  
      




\bibitem{16k}   Kolb, V.M. (2016)  Origins of Life: Chemical and Philosophical Approaches. Evol Biol. 43, 506–515.
{\textcolor[rgb]{0.00,0.00,0.84}{https://doi.org/10.1007/s11692-015-9361-4}} 
  


 



 \bibitem{8cc}     Krämer,  J.   et al.  (2022)  Molecular probes, chemosensors, and nanosensors for optical
detection of biorelevant molecules and ions in aqueous media and
biofluids.    Chem. Rev.            22(3), 3459–3636.
{\textcolor[rgb]{0.00,0.00,0.84}{https://doi.org/10.1021/acs.chemrev.1c00746}} 
  

 \bibitem{16kk}
Kristin, H. et al.  (2005) Sensitivity of cell-based biosensors to environmental variables,
Biosensors and Bioelectronics.  20(7), 1397-1406.
{\textcolor[rgb]{0.00,0.00,0.84}{https://doi.org/10.1016/j.bios.2004.06.007}} 






 \bibitem{19ll}             Lehn, J-M.	  (2007)        From supramolecular chemistry towards constitutional dynamic chemistry and adaptive chemistry                   Chem. Soc. Rev.  36, 151-160.
 {\textcolor[rgb]{0.00,0.00,0.84}{https://doi.org/10.1039/B616752G}}  
	

 \bibitem{19lll}    Liang,  W.  et al.  (2016) Identity recognition using biological electroencephalogram sensors, Journal of Sensors. 1-10. 
{\textcolor[rgb]{0.00,0.00,0.84}{https://doi.org/10.1155/2016/1831742}} 







\bibitem{20mc}
Malaterre.  C. and  Chartier, J.F.  (2021) Beyond categorical definitions of life: a data-driven approach to assessing lifeness. Synthese. 198, 4543–4572. 
{\textcolor[rgb]{0.00,0.00,0.84}{https://doi.org/10.1007/s11229-019-02356-w}} 






\bibitem{20m}     Mallatt,  J.  et al.   (2021)   Debunking a myth: plant consciousness.  Protoplasma.  258, 459–476.
{\textcolor[rgb]{0.00,0.00,0.84}{https://doi: 10.1007/s00709-020-01579-w}} 

\bibitem{20ma} 
Malaterre, C. (2010)  Lifeness signatures and the roots of the tree of life. Biol Philos 25, 643–658. 
{\textcolor[rgb]{0.00,0.00,0.84}{https://doi.org/10.1007/s10539-010-9220-8}}  


\bibitem{21mm}    Malaterre,  C.  et al.  (2022)
The origin of life: what is the question?
Astrobiology. 22, (7)851-862.
{\textcolor[rgb]{0.00,0.00,0.84}{http://doi.org/10.1089/ast.2021.0162}}  





\bibitem{20mm}  Merindol,  R. and  Walther,  A. (2017) Materials learning from life: Concepts for active, adaptive, and autonomous molecular systems. Chem. Soc. Rev.  46, 55-88. 
{\textcolor[rgb]{0.00,0.00,0.84}{https://doi.org/10.1039/C6CS00738D}}  


\bibitem{20mo} 
Monnard PA, Walde P.    (2015) Current Ideas about Prebiological Compartmentalization. Life (Basel).  10;5(2):1239-63. 
{\textcolor[rgb]{0.00,0.00,0.84}{https://doi: 10.3390/life5021239.}}  




\bibitem{20mv}
  Muchowska, K.B.  et al. (2022)  Nonenzymatic Metabolic Reactions and Life’s Origins.
Chemical Reviews. 120(15),  7708-7744.
{\textcolor[rgb]{0.00,0.00,0.84}{https://doi.org/10.1021/acs.chemrev.0c00191}}  
   





  
 \bibitem{24n}
Nader S, Sebastianelli L, Mansy SS.   (2022) Protometabolism as out-of-equilibrium chemistry. Philos Trans A Math Phys Eng Sci.  380(2227) 1-13.
{\textcolor[rgb]{0.00,0.00,0.84}{https://doi: 10.1098/rsta.2020.0423.}} 



 \bibitem{23pa}    Pascal, R. and Pross, A.  (2022) On the Chemical Origin of Biological Cognition. Life.  12(12), 1-12. 
{\textcolor[rgb]{0.00,0.00,0.84}{https://doi.org/10.3390/life12122016}}  
    
    


 \bibitem{23paa}   Pascal,  R.  et al.   (2013) Towards an evolutionary theory of the origin. of life based on kinetics and thermodynamics.
Open Biol.  3, 130-156. 
{\textcolor[rgb]{0.00,0.00,0.84}{https://doi.org/10.1098/rsob.130156}}  
    


\bibitem{23paaa}   Pascal, R. and Pross, A.  (2016) The logic of life. Origins of life and evolution of biospheres.  46, 507 -513.
{\textcolor[rgb]{0.00,0.00,0.84}{https://doi.org/10.1007/s11084-016-9494-1}}  
    





\bibitem{23pap}    Pascal, R. and Pross, A.  (2019) Chemistry’s kinetic dimension and the physical basis for life. Journal of Systems Chemistry. 7(1), 1-8. 
{\textcolor[rgb]{0.00,0.00,0.84}{https://hal.science/hal-02421013}}  
       
   




\bibitem{23po}    
 Pavlinova, P.  et al.  (2022)  Abiogenesis through gradual evolution of autocatalysis
into template-based replication.  FEBS letters. 597(3), 344-379.
{\textcolor[rgb]{0.00,0.00,0.84}{https://doi.org/10.1002/1873-3468.14507}}  
       
   


\bibitem{23p}        Pross, A and  Pascal, R. (2017)  How and why kinetics, thermodynamics, and chemistry induce the logic of biological evolution.
   Beilstein J Org Chem. 13, 665-674.
{\textcolor[rgb]{0.00,0.00,0.84}{https:/doi: 10.3762/bjoc.13.66. PMID: 28487761}}  
       
   
    

  \bibitem{23pp}    Pross, A. (2021) How was nature able to discover its own laws-twice?   Life.  12;11(7), 1-8.
{\textcolor[rgb]{0.00,0.00,0.84}{https://doi: 10.3390/life11070679}}  
  




\bibitem{24shap1} Shapiro,  J.A. (2007)   Bacteria are small but not stupid: cognition, natural genetic engineering and socio-bacteriology. Stud Hist Philos Biol Biomed Sci. 38(4), 807-819. {\textcolor[rgb]{0.00,0.00,0.84}{https://doi: 10.1016/j.shpsc.2007.09.010}} 



\bibitem{24shap}  Shapiro,  J.A. (2021) All living cells are cognitive,  Biochemical and Biophysical Research Communications.  564,  134-149.
{\textcolor[rgb]{0.00,0.00,0.84}{https://doi.org/10.1016/j.bbrc.2020.08.120}} 







   \bibitem{24st}       Stewart, J.E.  (2019) The origins of life: The Managed-Metabolism Hypothesis. Found Sci. 24, 171–195. 
{\textcolor[rgb]{0.00,0.00,0.84}{https://doi.org/10.1007/s10699-018-9563-1}} 


\bibitem{24ss}     Slootbeek, A.D.  et al.  (2022)  Growth, replication and division enable evolution of coacervate protocells. Chem. Commun. 58, 11183–11200.
{\textcolor[rgb]{0.00,0.00,0.84}{https://doi.org/10.1002/adbi.202200180}} 


 
\bibitem{8ccc}
 Somarakis, C. et al. (2016)
Interplays Between Systemic Risk and Network Topology in Consensus. 
 49(22)
 333-338.
{\textcolor[rgb]{0.00,0.00,0.84}{https://doi.org/10.1016/j.ifacol.2016.10.419.
}}  
 


  \bibitem{31t}   Trefil, J. et al.   (2009)  The origin of life.    American Scientist.    97(3), 206-214.
{\textcolor[rgb]{0.00,0.00,0.84}{https://doi:10.1511/2009.78.206}}  
 





 \bibitem{33w}   Whitesides,  G.M. and  Grzybowski, B. (2002) Self-assembly at all scales.  Science.  295, 2418–2421.
{\textcolor[rgb]{0.00,0.00,0.84}{https://doi:10.1126/science.1070821}}  

 
 \bibitem{33ww}   Winterbach, W., Mieghem, P.V., Reinders, M. et al. Topology of molecular interaction networks. BMC Syst Biol 7, 90 (2013).  
{\textcolor[rgb]{0.00,0.00,0.84}{https://doi.org/10.1186/1752-0509-7-90}} 
 


\bibitem{33}   Ziemann, A.E.  et al.  (2009) The amygdala is a chemosensor that detects carbon dioxide and acidosis to elicit fear behavior.  Cell.  25;139(5), 1012-1021.
{\textcolor[rgb]{0.00,0.00,0.84}{https://doi: 10.1016/j.cell.2009.10.029}} 


\end{thebibliography}
\end{document}